\documentclass[11pt]{article}
\pdfoutput=1
\usepackage{cite}
\usepackage{amsmath,bbm}
\usepackage{amssymb}
\usepackage{graphicx}
\usepackage[latin1]{inputenc}
\usepackage{mathrsfs}
\usepackage[dvipsnames]{xcolor}
\usepackage[footnotesize]{caption}
\usepackage{xcolor}
\usepackage{dsfont}
\usepackage{hyperref}
\usepackage[hmarginratio=1:1,top=32mm,columnsep=20pt]{geometry}
\usepackage{paralist}
\usepackage{abstract}
\usepackage{upgreek}

    \makeatletter
    \let\@fnsymbol\@arabic
    \makeatother

\newcommand{\bea}{\begin{eqnarray}}
\newcommand{\eea}{\end{eqnarray}}
\newcommand{\be}{\begin{equation}}
\newcommand{\ee}{\end{equation}}
\newcommand{\ba}{\begin{array}}
\newcommand{\ea}{\end{array}}

\def\gsim{\mathrel{\rlap{\lower4pt\hbox{\hskip1pt$\sim$}}
    \raise1pt\hbox{$>$}}}

\title{\vspace{-15mm}
	\fontsize{16pt}{10pt}\selectfont
	\textbf{Displaced vertex searches for sterile neutrinos \\ \medskip at future lepton colliders}
	}	
\author{%
	\large
	\textsc{Stefan~Antusch$^{\star \dagger}$\footnote{E-mail: \texttt{stefan.antusch@unibas.ch}},  Eros~Cazzato$^\star$\footnote{E-mail: \texttt{e.cazzato@unibas.ch}}, Oliver~Fischer$^\star$\footnote{E-mail: \texttt{oliver.fischer@unibas.ch}}}\\[10pt]
	\normalsize	$^\star$ Department of Physics, University of Basel, \\ 
	\normalsize 	Klingelbergstr.\ 82, CH-4056 Basel, Switzerland\\[5pt]
	\normalsize	$^\dagger$ Max-Planck-Institut f\"ur Physik (Werner-Heisenberg-Institut),\\
	\normalsize	F\"ohringer Ring 6, D-80805 M\"unchen, Germany
	\vspace{-5mm}
	}
\date{}

\begin{document}

\maketitle

\begin{abstract}
\noindent 
We investigate the sensitivity of future lepton colliders to displaced vertices from the decays of long-lived heavy (almost sterile) neutrinos with electroweak scale masses and detectable time of flight. As future lepton colliders we consider the FCC-ee, the CEPC, and the ILC, searching at the $Z$-pole and at the center-of-mass energies of 240, 350 and 500 GeV. For a realistic discussion of the detector response to the displaced vertex signal and the Standard Model background we consider the ILC's Silicon Detector (SiD) as benchmark for the future lepton collider detectors. We find that displaced vertices constitute a powerful search channel for sterile neutrinos, sensitive to squared active-sterile mixing angles as small as $10^{-11}$.

\end{abstract}

\section{Introduction}
Neutrino oscillation experiments have provided us with convincing evidence that at least two of the light neutrinos are indeed massive. 
The absolute mass scale of the light neutrino masses is bounded to lie below about $0.2$ eV from neutrinoless double beta decay experiments and cosmological constraints, see for instance ref.~\cite{Gariazzo:2015rra,deGouvea:2015euy} for recent reviews. 

An efficient and elegant extension of the Standard Model (SM), that aims at generating the light neutrinos' masses, is given by adding sterile (``right-handed'') neutrinos to its field content (see e.g.\ ref.\ \cite{Abazajian:2012ys} and references therein). The sterile neutrinos can have a so-called Majorana mass as well as Yukawa couplings to the three active neutrinos and to the Higgs doublet.
When the electroweak symmetry is broken, the sterile and active neutrinos mix, which yields light and heavy mass eigenstates that are each subject to a number of experimental constraints. 

The na\"ive one-family type I seesaw relation, given by $m_\nu \approx y^2\, v^2_\mathrm{EW}/M$, imposes either tiny neutrino Yukawa couplings $y$ (for Majorana masses $M$ on the electroweak scale), or Majorana masses around the Grand Unification scale (for Yukawa couplings of order one), such that an observation of this kind of heavy neutrino at colliders is not very promising.
This relation does not hold for seesaw scenarios with two or more sterile neutrinos with a protective symmetry, e.g.\ a ``lepton-number-like'' symmetry. In those scenarios, no constraints on the neutrino Yukawa couplings and the Majorana masses arise from the light neutrinos' mass scale (see e.g.\ refs. \cite{seesaw_low}). In this case the Yukawa couplings (or, alternatively, the active-sterile mixing angles) are theoretically unsuppressed, which in principle allows for effects to be searched for at particle colliders. 
A very interesting effect arises from heavy neutrinos with masses below the $W$ boson mass and with very small mixings. Such heavy neutrinos have suppressed couplings to the $W$ and $Z$ as well as to the Higgs boson $h$, which leads to a long enough lifetime for a potentially visible displacement from the interaction point. Via virtual $W,Z$ and $h$ they decay into the kinematically available SM particles.
This effect of a secondary vertex from the decays of the heavy neutrino, that is displaced from the primary vertex, yields an exotic signature and constitutes a powerful search channel for heavy neutrinos at particle colliders. This type of signature is being looked for by the ATLAS and CMS collaborations in their searches  for long-lived neutral particles \cite{Aad:2014yea,CMS:2014wda}.

The sensitivity of the LHC to heavy neutrinos via displaced vertex searches has been studied in refs.\ \cite{ConstraintsLHC}, which derived a constraint on the squared active-sterile mixing angle $| \theta |^2 < 10^{-5}$ for heavy neutrino masses $\sim 20$ GeV, and prospects for a sensitivity of $| \theta |^2 \sim 10^{-7}$ for 300 fb$^{-1}$. 
Estimates for the sensitivity reach of long-lived heavy neutrinos via displaced vertices have been made for the $Z$-pole run of the Future Circular electron positron Collider (FCC-ee) in ref.\ \cite{Blondel:2014bra}, showing a remarkable improvement over the LHC reach, and the results from the DELPHI collaboration \cite{Abreu:1996pa} at the Large Electron Positron collider (LEP). 

In this work, we deepen and extend the study in \cite{Blondel:2014bra}, where a first look at the sensitivity of displaced vertex searches for sterile neutrinos at the FCC-ee was taken. 
In particular, we put emphasis on the response of the detector to the heavy neutrino signal and the conceivable SM backgrounds.
As a benchmark detector for the future lepton colliders we consider the Silicon Detector (SiD), designed for the International Linear Collider (ILC). 
We derive realistic estimates for the sensitivities of the FCC-ee, the Circular Electron Positron Collider (CEPC) and the ILC.
These future colliders each have their unique physics program, defined by a target luminosity at the $Z$ pole, the Higgs threshold run\footnote{The FCC-ee presently considers the Higgs run at 240 GeV.} at 250 GeV, the top threshold scan at 350 GeV and for the ILC also 500 GeV.
Our theoretical framework is given by the Symmetry Protected Seesaw Scenario (SPSS) (see ref.\ \cite{Antusch:2015mia}), wherein the heavy neutrino masses and the active-sterile neutrino mixings are not subject to the na\"ive type I seesaw relation and thus not constrained by the light neutrinos' mass scale.

The paper is structured as follows. In section \ref{sec:model} we review the Symmetry Protected Seesaw Scenario. We discuss the production of heavy neutrinos in leptonic collisions and the lifetime of the produced heavy neutrinos before their decay in section \ref{sec:theory}. The future experiments are defined in section \ref{sec:displacement}, the response of the SiD is discussed, and the results of the analysis are presented. We summarise and conclude in section \ref{sec:conclusions}.

\section{The symmetry protected seesaw scenario}
\label{sec:model}
Sterile (or right-handed) neutrinos can have Majorana masses around the electroweak (EW) scale and unsuppressed active-sterile mixings, when they are subject to a ``lepton-number-like'' symmetry. 
The relevant features of seesaw models with this kind of protective symmetry, cf.\  refs.\ \cite{seesaw_low} for models with similar structures, may be represented by the benchmark model that was introduced in \cite{Antusch:2015mia} and which we refer to as the Symmetry Protected Seesaw Scenario (SPSS).
The SPSS considers a pair of sterile neutrinos $N_R^I$ $(I=1,2)$ and a suitable ``lepton-number-like'' symmetry, where $N_R^1$ ($N_R^2)$ has the same (opposite) charge as the left-handed $SU(2)_L$ doublets $L^\alpha$ $(\alpha=e,\mu,\tau)$. 
Light neutrino masses and other lepton-number-violating effects can be introduced by a small deviation from the exact symmetry limit.
The Lagrangian density of the SPSS, in the symmetric limit, is given by
\be
\mathscr{L} \supset \mathscr{L}_\mathrm{SM} -  \overline{N_R^1} M N^{2\,c}_R - y_{\nu_{\alpha}}\overline{N_{R}^1} \widetilde \phi^\dagger \, L^\alpha+\mathrm{H.c.}\;,
\label{eq:lagrange}
\ee
where $\mathscr{L}_\mathrm{SM}$ contains the usual SM field content and with $L^\alpha$ and $\phi$ being the lepton and Higgs doublets, respectively. The $y_{\nu_{\alpha}}$ are the complex-valued neutrino Yukawa couplings and the Majorana mass $M$ can be chosen real without loss of generality. 
We note that the SPSS allows for additional sterile neutrinos, provided their mixings with the other neutrinos are negligible, or their masses are very large, such that their effects decouple. This is a minimal framework that can explain the two observed mass squared differences of the light neutrinos and features four independent parameters relevant for collider experiments, namely the three $y_{\nu_{\alpha}}$ and $M$.

From eq.~(\ref{eq:lagrange}) we can derive the mass matrix ${\cal M}$ of the neutral fermions, which can be diagonalised with the unitary leptonic mixing matrix $U$ (a parametrization to $\mathcal{O}(\theta^2)$ can be found for instance in ref.\ \cite{Antusch:2015mia}):
\be
U^T\, {\cal M}\, U = \text{Diag}\left(0,0,0,M,M\right)\,.
\label{eq:diagonalisation}
\ee
The mass eigenstates are the three light neutrinos $\nu_i$ $(i=1,2,3)$, which are massless in the symmetric limit, and two heavy neutrinos $N_j$ $(j=1,2)$ with degenerate mass eigenvalues.
The mixing of the active and sterile neutrinos can be quantified by the mixing angles and their magnitude:
\be
\theta_\alpha = \frac{y_{\nu_\alpha}^{*}}{\sqrt{2}}\frac{v_\mathrm{EW}}{M}\,, \qquad |\theta|^2 := \sum_{\alpha} |\theta_\alpha|^2\,.
\label{def:thetaa}
\ee
Due to the mixing between the active and sterile neutrinos, the light and heavy neutrino mass eigenstates interact with the weak gauge bosons.

The present constraints from past and ongoing experiments and the sensitivities of future lepton colliders to the heavy neutrinos for the SPSS have been presented and discussed in \cite{Antusch:2015mia,Antusch:2014woa,Antusch:2015gjw}.
Further observable features of models with right-handed neutrinos have been investigated with respect to collider phenomenology in \cite{ConstraintsSterile}.

\section{Vertex displacement of heavy neutrinos}
\label{sec:theory}
\begin{figure}
\begin{minipage}{0.49\textwidth}
\centering
\includegraphics[scale=0.48]{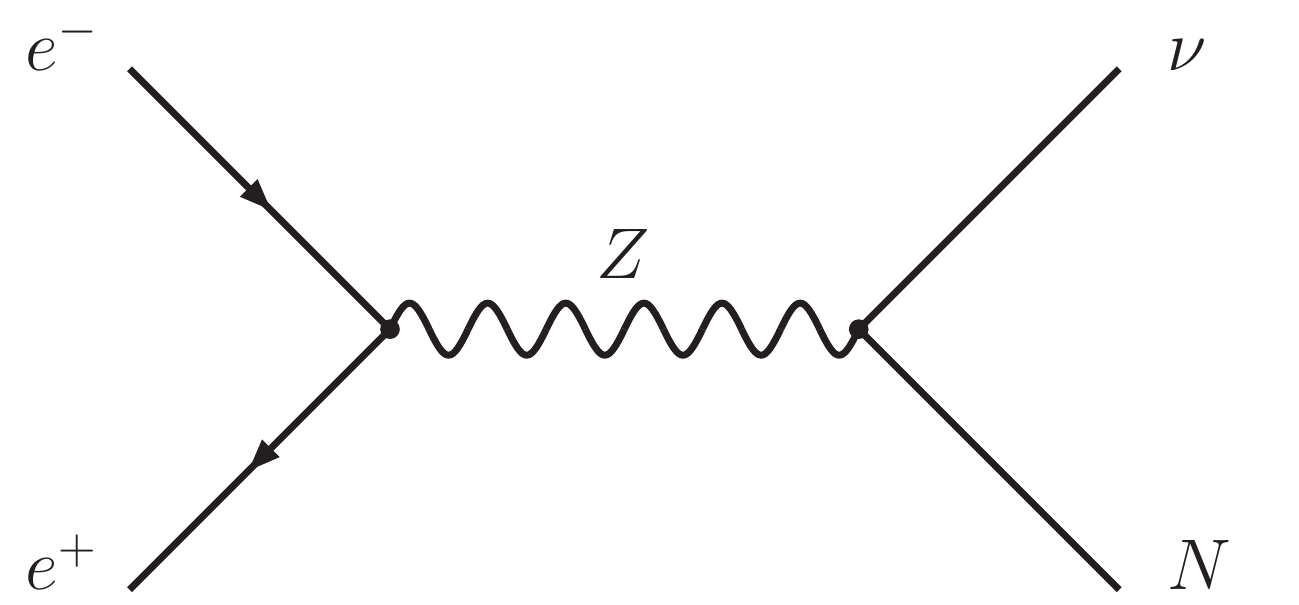}
\end{minipage}
\begin{minipage}{0.49\textwidth}
\centering
\includegraphics[scale=0.53]{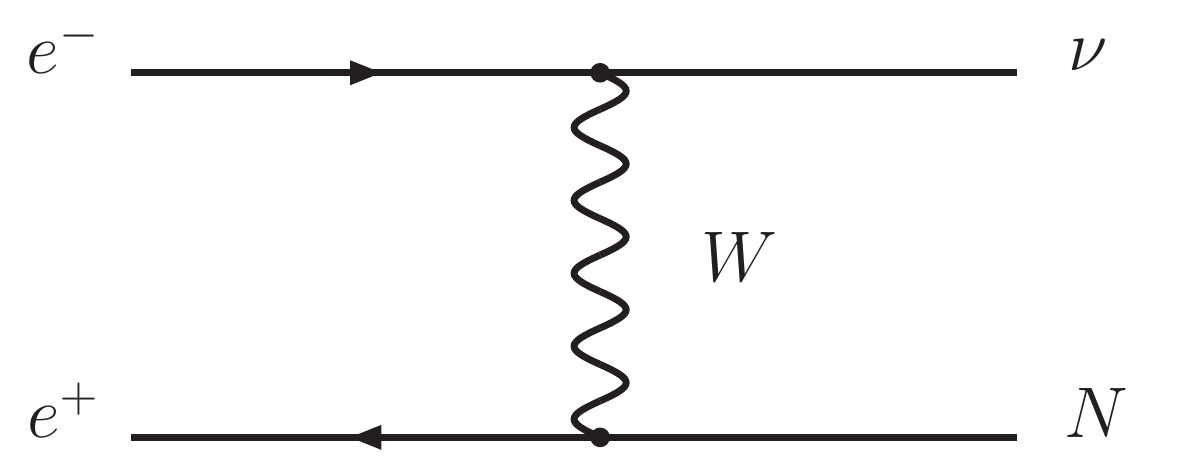}
\end{minipage}
\caption{Feynman diagrams contributing to the cross section for heavy neutrino production.}\label{fig:feyn}
\end{figure}
\begin{figure}
\begin{center}
\includegraphics[width=0.47\textwidth]{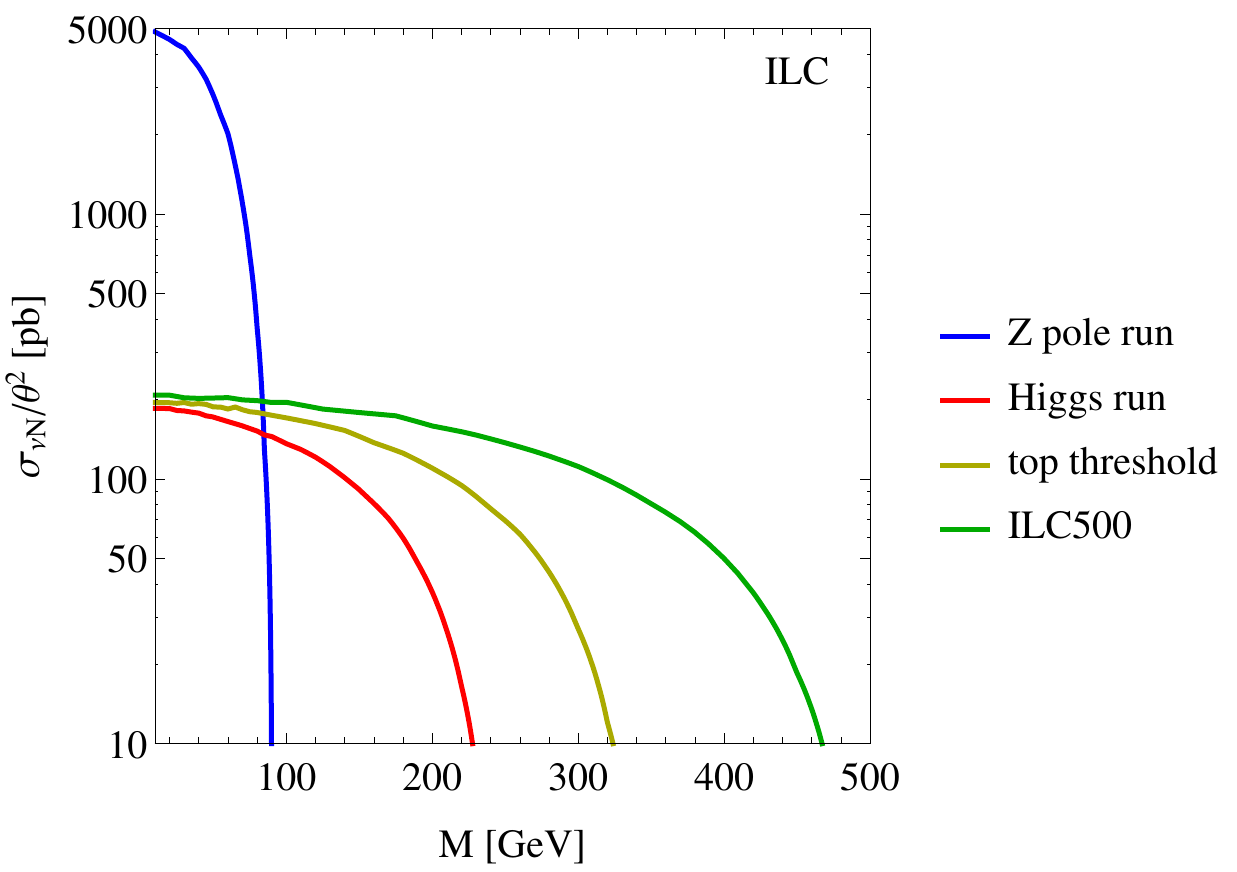}
\includegraphics[width=0.346\textwidth]{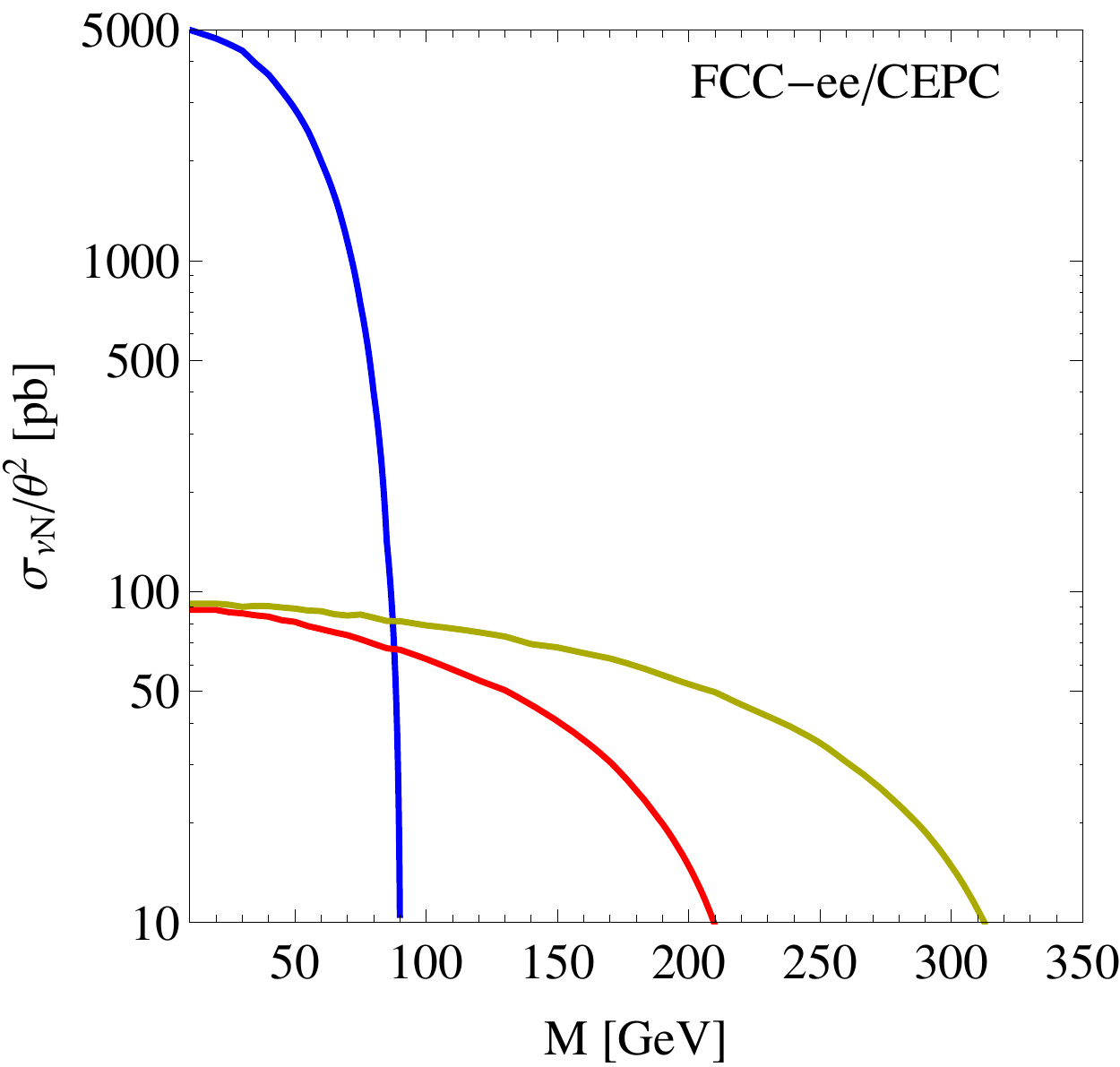}
\end{center}
\caption{Production cross section for heavy neutrinos at the FCC-ee, CEPC and ILC at different center-of-mass energies, divided by the square of the active-sterile neutrino mixing angle. Initial state radiation is included for both plots and for the ILC a (L,R) polarisation of (80\%, 30\%) and beamstrahlung are also included.}
\label{fig:Nproduction}
\end{figure}

In this section we introduce the preliminaries for the search for long-lived heavy neutrinos via displaced vertices.  
In this line we present the production mechanism for heavy neutrinos and the corresponding cross sections at lepton colliders for various center-of-mass energies. 
After their production a long-lived heavy neutrino can travel a finite distance before it decays, which is a stochastic process and follows an exponential probability distribution.
We quantify the number of heavy neutrinos that can be expected with a specific displacement.

The mechanisms of heavy neutrino production at $e^+e^-$-colliders are mediated by the weak gauge bosons and depicted by the Feynman diagrams in fig.\ \ref{fig:feyn}.
We define the heavy neutrino production cross section, to leading order in the small active-sterile mixing, by
\be
\sigma_{\nu N} = \sum_{i,j} \sigma(e^+ e^- \to N_j\,\nu_i)\,,
\label{eq:sigmanuN}
\ee
where we sum over all the light neutrinos ($i = 1,2,3$) and the two heavy neutrinos ($j=1,2$), and the cross section is a function of the center-of-mass energy $\sqrt{s}$.
We implemented the SPSS via Feynrules \cite{Alloul:2013bka} into the Monte Carlo event generator WHIZARD 2.2.7 \cite{Kilian:2007gr,Moretti:2001zz} and evaluated the heavy neutrino production cross section, including initial state radiation for all colliders, and lepton beam polarisation for the ILC, for the center-of-mass energies 90, 250, 350 and 500 GeV, as shown in fig.\ \ref{fig:Nproduction}.
At the $Z$-pole, the production of heavy neutrinos via the $s$-channel $Z$ boson is dominant, which is sensitive to all neutrino Yukawa couplings $|y_{\nu_\alpha}|$ $(\,\alpha=e,\mu,\tau)$. However, for larger center-of-mass energies, the dominant contribution to the production cross section of the heavy neutrinos comes from the t-channel exchange of a $W$ boson, which is only sensitive to $|y_{\nu_e}|$.

\begin{figure}
\begin{center}
\includegraphics[width=0.56\textwidth]{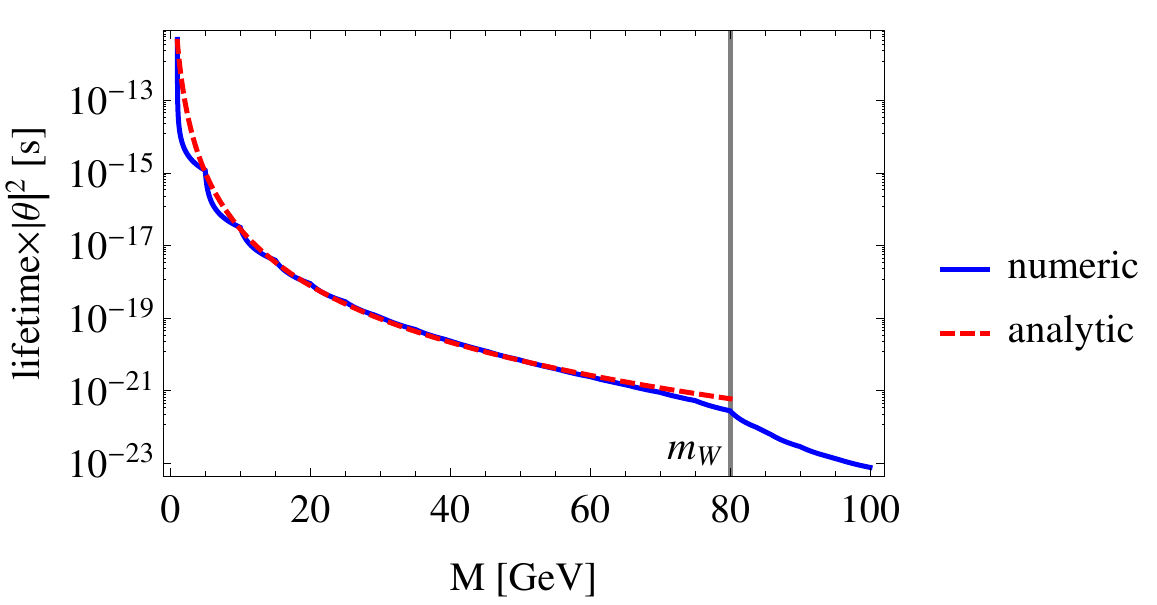}
\end{center}
\caption{Lifetime (in seconds) of a heavy neutrino as a function of its mass, divided by the active-sterile mixing $|\theta|^2$. The analytic line is valid only for $M<m_W$ and was evaluated with eq.\ \eqref{eq:lifetime}, while the numerical evaluation of the lifetime was done with WHIZARD.}
\label{fig:lifetime}
\end{figure}

The heavy neutrinos decay into SM particles and their lifetime $\uptau$ is given by the inverse of the decay width $\Gamma_N$. For heavy neutrinos lighter than the $W$ boson mass, their decays, which occur via off-shell gauge and Higgs bosons, are suppressed. Furthermore, for small active-sterile mixing angles the $|\theta|^2$ dependency on $\Gamma_N$ can render the heavy neutrinos long-lived compared to SM particles. We evaluate the proper lifetimes from the decay widths for heavy neutrino masses larger than 1 GeV with WHIZARD.
In fig.\ \ref{fig:lifetime} we show the resulting proper lifetime as a function of the heavy neutrino mass, together with the analytical formula from ref.\ \cite{Gronau:1984ct} 
\begin{equation}
\uptau_e = \frac{4.15 \times 10^{-12}}{|\theta_e|^{2}} \left(\frac{\rm GeV}{M}\right)^{5.17}\, {\rm s}\,, \label{eq:lifetime}
\end{equation}
where only the neutrino coupling to the electron flavour is considered. 
The heavy neutrino lifetime in eq.~\eqref{eq:lifetime} is valid for $M<m_W$, and the analytical formula is in good agreement with the obtained numerical result. 
As one can see in the figure, the lifetime is reduced as additional decay channels open up, especially when the heavy neutrino mass exceeds $m_W,\,m_Z$ or $m_h$ as the decay channels via on-shell gauge or Higgs bosons become efficient.

Depending on the lifetime of the heavy neutrino, it can travel a finite distance after its production at a particle collider before it decays.
The heavy neutrino lifetime in the laboratory frame is related to the proper lifetime by
\begin{equation}
\uptau^{\rm lab} = \gamma\,\uptau\,,
\end{equation}
with the Lorentz factor $\gamma = \sqrt{M^2+ |\vec{p}_N|^2}/M$, where $\vec{p}_N$ is the three momentum of the heavy neutrino. 
Due to the production process of the heavy neutrino being a 2$\to$2 process with one massive particle in the final state, the magnitude of the three momentum can be expressed as
\begin{equation}
|\vec{p_N}| = {1\over2}\left( \sqrt{s}-\frac{M^2}{\sqrt{s}}\right)\,,
\end{equation}
with $\sqrt{s}$ being the center-of-mass energy. 
The decay of a long-lived heavy neutrino is a stochastic process and follows an exponential probability distribution. Its probability to decay with a displacement 
$x$ from the primary vertex with $x_1 \leq x \leq x_2$ (where $x_1$ and $x_2$ are an inner and outer boundary) is given by
\begin{equation}
P(x_1,x_2)= \int_{t_1}^{t_2} \frac{1}{\uptau^{\rm lab}}\, \exp\left(-\frac{t}{\uptau^{\rm lab}}\right)dt\,,
\label{eq:probability}
\end{equation}
with $t_i=x_i/|\vec{v}|$ distance  and the velocity $|\vec{v}| = \frac{|\vec{p_N}|}{E_N}$ (in natural units).
Combining production and decay of heavy neutrinos at lepton colliders yields the expected number of heavy neutrinos, which are produced at the interaction point and decay with a displacement of at least $x_1$ and at most $x_2$:
\begin{equation}
N(x_1,x_2,\sqrt{s},{\cal L}) = P(x_1,x_2)\, \sigma_{\nu N}(\sqrt{s})\, {\cal L}\,,
\label{eq:nevents}
\end{equation}
with the integrated luminosity $\cal{L}$ and the heavy neutrino production cross section from eq.\ \eqref{eq:sigmanuN} and in fig.\ \ref{fig:Nproduction}. 
Contextualising the formula eq.\ \eqref{eq:nevents} with the considered future lepton collider experiments is the subject of the next section.

\section{Displaced vertices at future lepton colliders}
\label{sec:displacement}
\begin{figure}[h]
\centering
\includegraphics[scale=1.1]{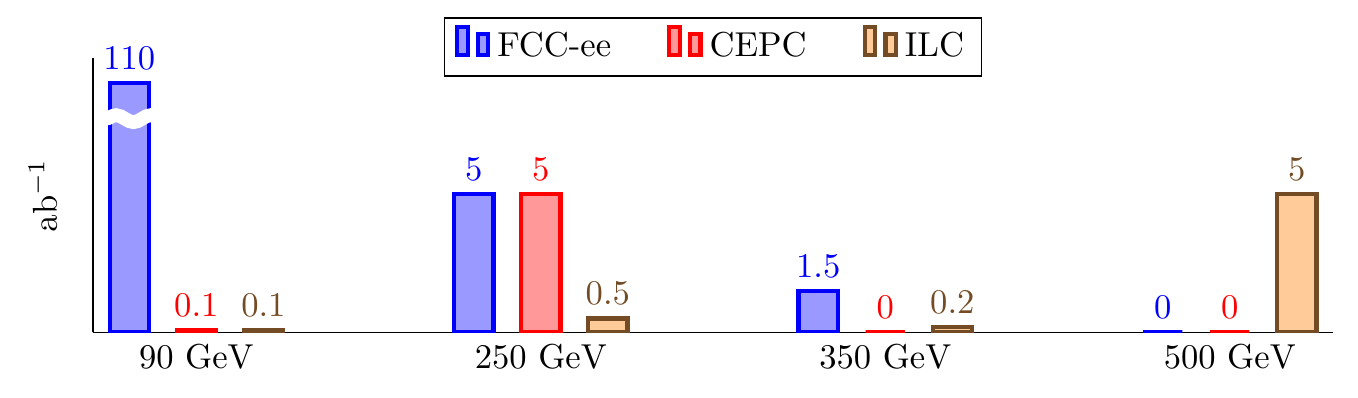}
\caption{Proposed modi operandi, defined by target integrated luminosities for each center-of-mass energy, for the considered future lepton colliders. 
For the FCC-ee \cite{Gomez-Ceballos:2013zzn} we use the product of the target instantaneous luminosities from \cite{FCCweb} (for two interaction points) and the envisaged run-times, and the Higgs run with a center-of-mass energy of 240 GeV.
For the CEPC we use the exemplary integrated luminosities from the preCDR \cite{preCDR}.
For the ILC \cite{Baer:2013cma} we consider the G-20 operation scenario from ref.\ \cite{Brau:2015ppa}, and we further include the Giga-Z operation.}
\label{fig:modioperandi}
\end{figure}

The search for sterile neutrinos via displaced vertices is considered at the planned future lepton colliders, the FCC-ee, the CEPC, and the ILC, each with its own physics program as shown in fig.~\ref{fig:modioperandi}. 
We note that we chose the operation scenario G-20 for the ILC because it considers the most integrated luminosity at 500 GeV, anticipating a more promising sensitivity. We furthermore add the Giga-Z operation, for which we reckon with 100 fb$^{-1}$ (resulting in $\sim 10^9$ $Z$ bosons at the $Z$ pole).

Apart from the explicit modus operandi of a future lepton collider, the search for heavy neutrinos via displaced vertices also depends on the detector layout and its performance parameters. 
In the following we assess the detectability of the signal and possible SM backgrounds. For definiteness we consider the ILC's Silicon Detector (SiD) \cite{Aihara:2009ad,Behnke:2013lya} as benchmark,
which is chosen as an example and can be expected to yield a performance that is comparable to other planned detectors, e.g.\ the ILD \cite{Behnke:2013lya,Abe:2010aa}. 

The SiD's integrated tracking system is developed for the particle flow algorithm, and it consists in a powerful silicon pixel vertex detector, silicon tracking, silicon-tungsten electromagnetic calorimetry (ECAL) and highly segmented hadronic calorimetry (HCAL). Furthermore, the detector layout incorporates a high-field solenoid, and an iron flux return that is instrumented as  muon identification system. 
The SiD geometry, separated into the barrel and the endcap, allows for a high level of hermeticity with uniform coverage and a transverse impact parameter resolution of $\sim 2\,\mu$m over the full solid angle.
In the following, we assume a spherical symmetry for the SiD, which is sufficient for our analysis, and use the radii of the individual detector components from the barrel part, which are summarised in tab.\ \ref{tab:SiD}.
\begin{table}
\begin{center}
\begin{tabular}{c|ccc}
\hline
(Barrel) &	  Inner radius &	Outer radius &	z extent \\
\hline\hline
Vertex detector	&  1.4	&	6.0		&	+/- 6.25 \\
Tracker	Silicon &  21.7	&	122.1	&	+/- 152.2 \\
ECAL	 		& 126.5&	140.9	&	+/- 176.5 \\
HCAL	 		& 141.7	&	249.3	&	+/- 301.8 \\
Solenoid	 	& 259.1&	339.2	&	+/- 298.3 \\
Flux return	& 340.2 & 	604.2	&	+/- 303.3 \\
\hline	
\end{tabular}
\end{center}
\caption{SiD barrel structure, radii in cm. Taken from ref.\ \cite{Behnke:2013lya}.}
\label{tab:SiD}
\end{table}

\subsection{Signal and background}
We call the SiD response to the heavy neutrino decay products the signal, and its response to SM processes we call the background. 

The possible final states for heavy neutrino mass $M<m_W$, with approximate branching ratios, are
\begin{equation}
\begin{array}{lrcrl}
{\rm Br}(N_i\to \nu\nu\nu) & \sim & 5\% & \text{invisible}\,, \\
{\rm Br}(N_i\to \nu\ell^+\ell^-) & \sim & 25\% & \text{leptonic}\,, \\
{\rm Br}(N_i\to \nu q \bar q) & \sim & 15\% & \text{hadronic}\,, \\
{\rm Br}(N_i\to \ell^\pm q^\prime \bar q) & \sim & 55\% & \text{semileptonic}\,.
\end{array}
\end{equation}
The branching ratios have a small dependency on the mass $M$, that will be neglected in the following.
The heavy neutrinos are produced together with a light neutrino, such that their decay products are always associated with missing momentum. The experimental signature that arises from the decay of a {\bf long-lived} heavy neutrino is given by {\bf exactly one} secondary vertex, from where all visible particles in the detector originate.

The striking feature of only one visible secondary vertex makes the experimental signature of heavy neutrino decays very  distinct from possible SM processes. The discussion of the backgrounds in the following is based on the simulation of ${\cal O}(10^7)$ SM events with WHIZARD \cite{Kilian:2007gr,Moretti:2001zz} that were reconstructed with DELPHES \cite{Cacciari:2011ma} using the DSiD detector card \cite{Potter:2016pgp}.
We consider SM processes with the following final states: $f\bar f,\, f\bar f \gamma,\, f\bar f \gamma \gamma,\, f\bar f \nu \nu,\,\nu\ell q \bar q^\prime$, with $f$ being a charged lepton or a quark, $\gamma$ a photon and $\nu$ a light neutrino.
From the considered final states, especially $f\bar f$ and events with neutrinos may give rise to a viable background in the following way:
\begin{itemize}
\item {\bf Loss of particles in the beam pipe:} Final state particles with a sufficiently small transverse momentum can remain inside the beam pipe and thus prevent detection. If one such particle recoils e.g.\ against an ISR photon it may get ``kicked'' into the detection volume, featuring typically a very small angle to the beam axis. This type of event could be vetoed against with hard gammas, or, similarly, with the angle between beam axis and visible particle. Furthermore, in this type of background the overall charge of the event may be measured as non-zero, which could provide the most powerful veto.
\item {\bf Miss-reconstructed events:} It is possible that a reconstruction algorithm does not identify a normally visible particle. These events can be vetoed against via energy deposits that are located in the detector region opposite the observed particle. Moreover, the overall charge can also be used as a veto.
\item {\bf Merging of secondary vertices:} When particles with finite lifetime are produced in pairs and decay sufficiently close to each other, such that their individual secondary vertices cannot be resolved from the tracking information, this can constitute a background. 
This implies that the particles have to be emitted in a very narrow solid angle, which necessitates the production of additional invisible particles, (i.e.\ light neutrinos) in order to balance the overall momentum. This removes the contribution from the $f\bar f$ events.
\end{itemize}
In the following, we assume that the above mentioned vetoes remove all the possible backgrounds from processes with one lost or miss-reconstructed particle. Although a loss of signal efficiency is to be expected by such vetoes, we do not consider this in the following. For a quantitative statement on the veto efficiency, a detailed analysis of this type of events after a full detector simulation is needed, which is beyond the scope of the present analysis.

\begin{figure}
\begin{minipage}{0.49\textwidth}
\centering
\includegraphics[width=0.4\textwidth]{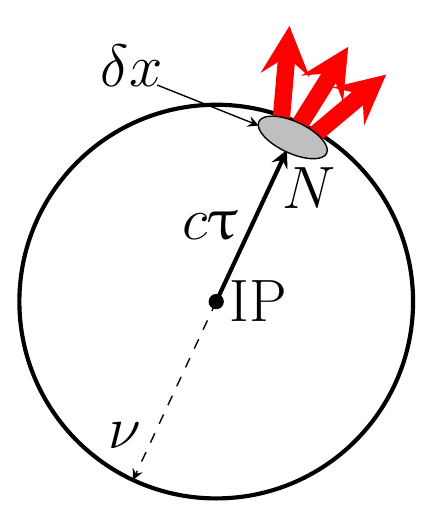}

{\bf signal}
\end{minipage}
\begin{minipage}{0.49\textwidth}
\centering
\includegraphics[width=0.4\textwidth]{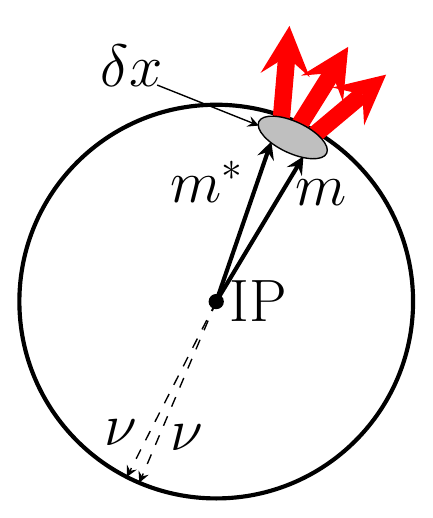}

{\bf background}
\end{minipage}
\caption{Schematic illustration of the {\bf signal}, that is given by the decays of a heavy neutrino at a distance $c\uptau$ from the interaction point. The SM {\bf background} is given by two light neutrinos ($\nu\nu$) and two long-lived mesons $m$ and $m^*$, which decay sufficiently close to each other, such that only one secondary vertex can be resolved. The detector resolution $\delta x$ is dependent on the detector component.}
\label{fig:signalbackground}
\end{figure}

\begin{figure}
\begin{center}
\includegraphics[width=0.4\textwidth]{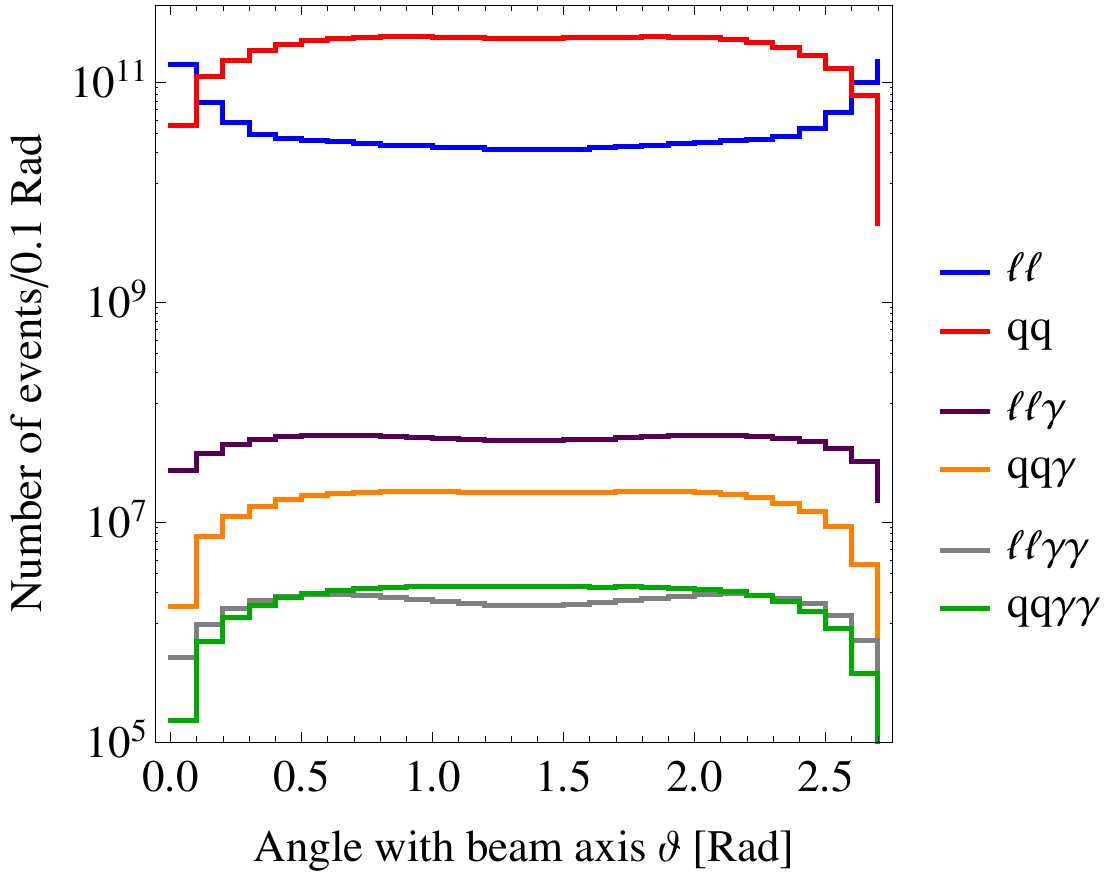}
\end{center}
\caption{Angular distribution of various SM processes for $\sqrt{s}=m_Z$ and an integrated luminosity of 100 ab$^{-1}$. The label $\ell\ell$ and $qq$ denotes the fermionic final states (after reconstruction) $e,\,\mu,\,\tau$ and $u,d,s,c,b$, respectively. The exponential increase in the number of $\ell\ell$ events close to the beam axis stems from the low-angle Bhabha scattering of electrons. The distributions at other center-of-mass energies are similar.}
\label{fig:thetadistribution}
\end{figure}

This promotes the merging of secondary vertices to the primary source of background for the displaced vertex searches. We show a schematic illustration of the displaced signal and backgrounds in fig.\ \ref{fig:signalbackground}. The probability of the decays of both SM particles occur within an unresolvable distance can be assessed with eq.~\eqref{eq:probability} and by taking into account the narrow solid angle. 
When isotropic emission of the fermionic final states is assumed, see fig.\ \ref{fig:thetadistribution}, the fraction of two mother particles that are emitted into a narrow solid angle can be estimated by $\Omega/4\pi$, with $\Omega = 2\pi \int^{\alpha}_0 \sin \theta d \theta$, $\alpha= \arcsin(\delta x/2x)$, where $\delta x$ is the spatial resolution of the detector and $x$ the distance from the IP (i.e. the displacement).

To assess the expected signal efficiencies we simulated $10^6$ events of semileptonic decays for $M=10$ and 40 GeV, respectively. The fast reconstruction with DSiD\footnote{We have adjusted the photon, electron, and muon isolation criteria to PTRatioMax = 0.12, 0.12, 0.25, respectively, and we set DeltaRMax = 0.5 for all three.} yields a signal efficiency of $\sim 80\%$ to find at least one jet for $M=10$ GeV. For $M=40$ GeV we find the efficiencies of $\sim99\%,\, \sim60\%$ and $\sim 20\%$ to find one lepton, one jet, and two jets, respectively. 
Since one visible object is sufficient for the identification of a displaced vertex, we assume that the signal can be observed with 100\% efficiency in the following.

\subsection{Detector response: the search for heavy neutrinos via displaced vertices}
\label{sec:SiD}
The vertex displacement $x$ is defined by the distance between the primary vertex where a mother particle (e.g.\ the heavy neutrino or a SM particle with finite lifetime) was produced, and the secondary vertex where the mother particle decays into a number of daughter particles.
Since the primary vertex is experimentally unknown, we consider the center of the interaction point instead and use its extension as uncertainty.

Depending on the displacement of the long-lived heavy neutrino, its decay can take place in any of the SiD's detector components. Therefore every component can be considered as an independent probe for displaced vertices with a well defined boundary given by its extension, see tab.\ \ref{tab:SiD}. 
In the following, we discuss the search for long-lived heavy neutrinos by investigating the individual response of the SiD components to their decays, and possible backgrounds.

\subsubsection{Inner region:}
\label{sec:innerregion}
We define the inner region as the volume that is enclosed by the vertex detector.
The vertex displacement $x$ can generally be inferred from the tracks of the decay products in the vertex detector and the tracker. 
The precision of $x$ is limited by the resolution of the tracker and the spatial extension of the Interaction point (IP).
At the ILC the IP has a vertical extension of $\sim 10$ nm and we assume a vertical extension of $\sim 250$ nm for the circular FCC-ee and CEPC for all the modi operandi.
The impact parameter resolution of the SiD in the transverse plane is $\sim 2\, \mu$m. We note that due to our assumption of a spherical detector geometry, this parameter is also valid in the longitudinal direction.
We therefore consider the resolution for displaced vertices $x_{\rm res}$, i.e. the minimum vertex displacement that is separable at $3\sigma$ from the IP, to be given by 6 $\mu$m for the ILC and 7 $\mu$m for the FCC-ee/CEPC.
We remark that this resolution is strictly valid for vertical displacements only. We note that for an accurate assessment of the resolution the entire geometry of the detector  and of the IP have to be taken into account, which, however, is beyond the scope of this paper.

\paragraph{Conventional search, $x < x_{\rm res}$:}
For heavy neutrino decays with $x$ smaller than $x_{\rm res}$, the vertex displacement cannot be used to distinguish between signal and background. 
This necessitates a conventional search, where the kinematic distributions are used to distinguish the heavy neutrino signal from the SM background.
For instance such a search for neutral heavy neutrinos produced in $Z$ decays was conducted by DELPHI at LEP I\ \cite{Abreu:1996pa}. The dominant SM background is given by four-fermion semileptonic and hadronic final states with missing momentum, namely $\ell^\pm \nu q^\prime \bar q$ and $f\bar f \nu\nu$, for $f$ being a charged SM fermion and $q=u,\,d,\,s,\,c,\,b$.
We assume that a conventional search for heavy neutrinos is possible as long as their decays have a displacement smaller than the outer radius of the tracker.

\paragraph{Search for displaced vertices, $x \geq x_{\rm res}$:}
The SM particles with lifetimes that can lead to a displaced secondary vertex, that occur dominantly in the inner region, are the $\pi^0$ meson ($c\uptau \sim$ 20 nm), the $\tau$ lepton ($c\uptau \sim$ 0.1 mm), and all the $D$ and $B$ mesons (with $c\uptau \sim$ 0.1 -- 0.5 mm).
Any of these particles can fake the heavy neutrino signal given that they result in only one secondary vertex, implying that they are accompanied by two light neutrinos.
In the inner region, two individual vertices cannot be separated when they are closer than 6 $\mu$m, which is set by the tracking resolution of the decay products at $3\sigma$, and not to be confounded with $x_{\rm res}$.

We estimate the production cross section of the neutral pions to be smaller than $\sigma(e^+ e^- \to q\bar q \nu \nu) \simeq 100$ fb (for $\sqrt{s}=m_Z$ and without polarisation), which sets an upper limit of $10^7$ events (at the FCC-ee for 110 ab$^{-1}$) among which some may fake the heavy neutrino signal. 
Most pions decay into two photons, which cannot be mistaken as a signal event. 
The fraction of events that lead to a ``signal-like'' final state, for instance $\nu \nu e^+ e^- \gamma$, is only ${\cal O}(10^{-7})$. Furthermore, demanding that the two secondary vertices cannot be distinguished from each other at 3 $\sigma$, i.e.\ that their respective vertices are at most 6 $\mu$m apart (see fig.~\ref{fig:signalbackground}), reduces the number of events by another factor of $10^{-3}$.

The number of ``signal-like'' background events from $\tau$ leptons is slightly larger than those of $D$ and $B$ mesons.
Considering the simultaneous decay of two $\tau$ leptons with cross section $\sigma(e^+ e^- \to \tau^+ \tau^- \nu \nu) \simeq 2$ fb at $\sqrt{s}=m_Z$ (no polarisation), we find that less than 1 event for 110 ab$^{-1}$ can be expected with a displacement of $x \geq 10\,\mu$m at the FCC-ee\footnote{This number is valid only for the FCC-ee. The CEPC, and also the ILC, with their smaller integrated luminosities, have much less than one expected event with vertex displacements as small as 6 $\mu$m.}.
We find that the backgrounds at higher center-of-mass energies are also effectively suppressed below one event for $x > 10\, \mu$m (by the requirement that the mesons or the $\tau$ leptons are emitted into a narrow solid angle).

Another background for the searches at the $Z$ pole may be the process $e^+e^-\to \tau^+\tau^- \gamma$ due to its large cross section of $\sim 1.6$ nb. When the $\gamma$ is hard and emitted in the direction of the beam pipe, it can escape detection. The condition of close-by decays of the two tau leptons within the inner region yields a suppression factor of $\sim 10^{-6}$, which leaves e.g.\ $\sim 10^5$ potential background events at the FCC-ee. We remark however that the invariant mass of the decay products may allow to discriminate against this and the other similar backgrounds, when the mass $M$ of the heavy neutrino is larger than the combined rest masses $\sim m_{m} + m_{m'}$ (e.g.\ $\sim 2m_\tau$) of the two decaying particles. This may allow to resolve vertex displacements closer to the IP than 10 $\mu$m for $M > 2 m_\tau$, while maintaining an almost background-free environment.

\subsubsection{Vertex detector and tracker}
The vertex detector is designed to detect the displaced vertices of heavy flavours for their efficient identification.
The highly efficient charged particle tracking allows to recognise and measure prompt tracks in conjunction with the ECAL.
The vertex displacement can be inferred from the impact parameter of the reconstructed tracks. We note at this point, that the resolution of the impact parameter degrades at the SiD when the heavy neutrino decays take place deep inside the tracker.  In this case not all the silicon layers would respond to the tracks from the secondary vertex, which would reduce the resulting impact parameter resolution. Because we expect all the future lepton colliders to have at least one detector with continuous tracking, which has a larger number of layers and thus might experience less degradation of the resolution, we ignore this effect in the following.

For heavy neutrino decays inside the vertex detector/tracker, all kinematic information on the decay products are available, in particular the vertex displacement. Moreover, since the heavy neutrinos are neutral, the displacement becomes directly visible by an appearing secondary vertex whence the decay products emerge.

SM particles which have a vertex displacement that results in their decays taking place dominantly within the vertex detector, are given by the $K_S$ ($c\uptau \sim$ 2.68 cm) meson and the $\Lambda$ baryon ($c\uptau \sim$ 7.89 cm).
We estimate that the condition of two SM particles with finite lifetime to be emitted into a narrow solid angle and to decay close to each other, to reduce the background by a factor $< 10^{-16}$ for both, the $K_S$ the $\Lambda$ baryon, such that those contributions are completely negligible.

\subsubsection{ECAL and HCAL} 
The calorimeter system has imaging capabilities that allow for efficient track-following, with a pixel size of $\sim 4 \mu$m for the ECAL and $\sim 1$ cm for the HCAL, which allows a correct association of energy clusters with tracks.

The heavy neutrino signal, with vertex displacements that result in decays in the calorimetric system, consists in one or more clusters of energy deposits, that should be connected and consistent with one secondary vertex. 
Defining features of this signal are the absence of tracks and significant momentum imbalance. 
Due to a lack of tracking information, the decay products may, however, be identified as electrically neutral particles.

The ECAL could record the leptonic part of the signal as a photon. The SM background for this signal contains at least one photon, and missing energy, for instance a pair of light neutrinos and a hard photon, $\nu\nu \gamma$. 
The production cross section for this background process 
is ${\cal O}($1) fb at $\sqrt{s}=m_Z$ for photon energies $\sim 10$ GeV and up to ${\cal O}(100$) fb at $\sqrt{s}=500$ GeV for the ILC. 
It may be possible to separate the signal efficiently from the background, but, to be conservative, we shall not consider the leptonic decays of the heavy neutrinos that take place inside the ECAL.

From the hadronic and semileptonic decays of the heavy neutrinos that take place in the HCAL, the decay products could be recorded as neutral hadrons. 
The SM particles that could fake a background are the $K_L$ ($c\uptau \sim$ 15.34 m). We estimate the production cross section to be smaller than the production cross section $q\bar q \nu \nu$, which is ${\cal O}(1)$ ab for $\sqrt{s}=m_Z$, and is $\sim$ 600 fb for $\sqrt{s}=$ 500 GeV.

Viable backgrounds can come from $\tau^+\tau^-\gamma$ and $\tau^+\tau^-\nu\nu$ events, where the tau leptons decay into two $K_L$.
If the tau leptons are collimated such that the two $K_L$ enter the HCAL with a maximum distance of at most 1 cm, such that the readout cannot yield an indication of the energy deposits being disconnected, this process may yield a fake signal event.
 The conservation of charge, however, leaves a charged lepton or a charged meson among the decay products of each tau lepton, which can be used as veto against such an event. Furthermore, the $\tau^+\tau^-\gamma$ events are very close to the beam axis, which will allow to suppress them against the signal distribution if necessary. We therefore assume that no background event remains.

\subsubsection{Muon identification system} 
\label{sec:mucal}
The muon detecting photomultipliers, intertwined with the steel layers of the solenoid flux return, identifies muons from the interaction point with high efficiency and rejects most of the remaining hadrons that spill over from the HCAL.
The muon selection combines the information from the tracker, the calorimetric system, and the muon detectors, to reconstruct the muon candidates.

A highly relativistic heavy neutrino will reach the outer radius of the flux return yoke in about 20 ns.
For its decays inside the flux return yoke, the resulting decay products interact with the scintillator bars. The ensuing photons are detected with the photomultipliers, and, due to the absence of information from the calorimeters and the tracker, they should not be identified as muon candidates.
Instead, they leave a number of hits in the photomultipliers for which there are no SM background processes from leptonic collisions.

The background in this case is given by cosmic ray muons, which can be rejected efficiently by correlating the corresponding hits with the beam collision time.
Further backgrounds are given by muons that are created in the interaction of the electron or positron beam with the beam-delivery system, and subsequently traverse the detector parallel to the beam line, also referred to as ``fliers''.
We assume, that all the visible heavy neutrino decay channels leave a measurable imprint in at least one layer of the photomultipliers.

\subsubsection{Combined response of the SiD}
\begin{figure}
\begin{minipage}{0.6\textwidth}
\begin{flushright}
\includegraphics[width=0.75\textwidth]{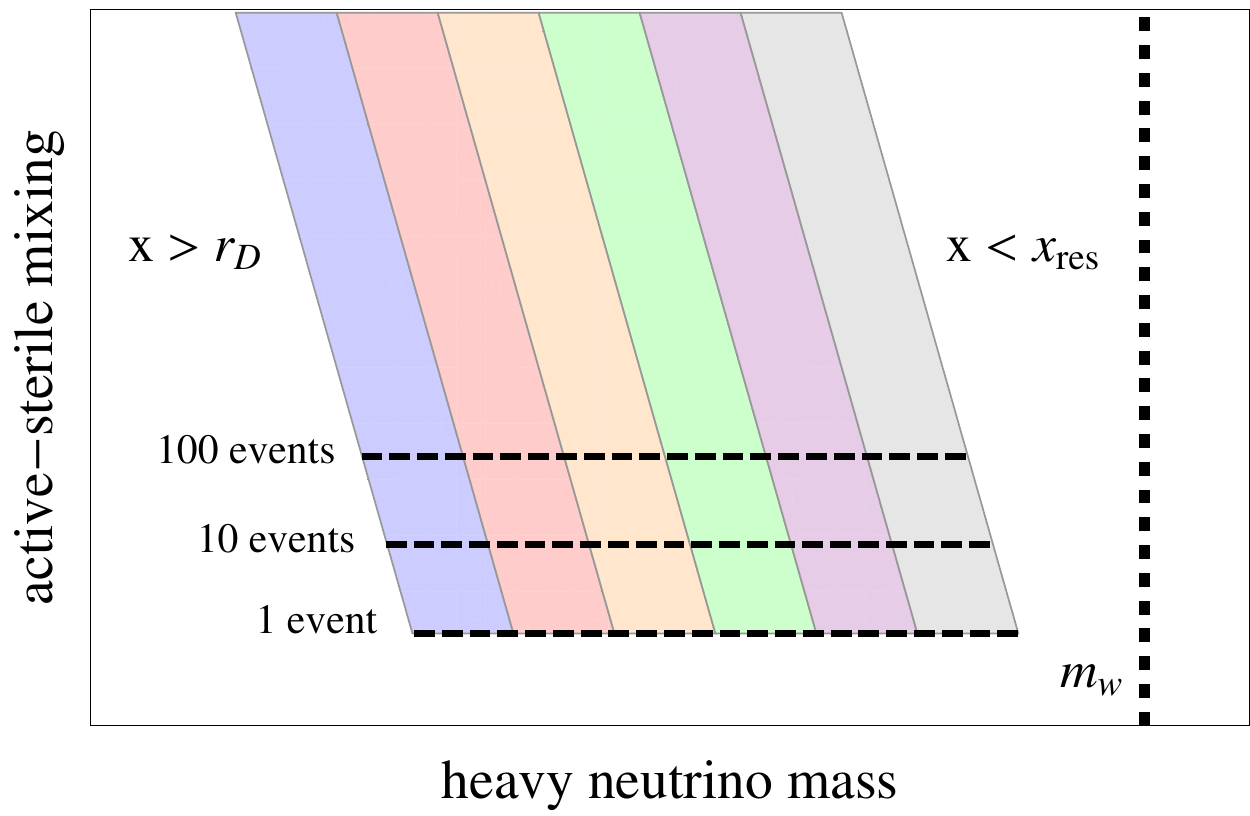}
\end{flushright}
\end{minipage}
\begin{minipage}{0.39\textwidth}
\includegraphics[width=0.45\textwidth]{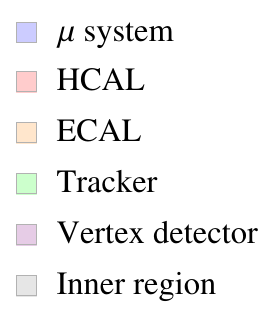}
\end{minipage}
\caption{Schematic illustration of the sensitivity of the different detector components to heavy neutrino decays as a function of the active-sterile mixing parameter and the heavy neutrino mass. The parameter $r_D$ is the outer radius of the muon system. We note that the sensitivities of the individual components are overlapping, such that it is not possible to assign one responsive detector component to one specific set of heavy neutrino parameters.}
\label{fig:schematic}
\end{figure}
In the following we discuss the sensitivities of displaced vertex searches for long-lived heavy neutrinos from the individual components of the SiD, and their combination.

Every individual detector component is (in principle) sensitive to the signal from long-lived heavy neutrinos, and a signal significance $\mathcal S$ can be established via 
\begin{equation}
\mathcal{S} = \frac{N_S}{\sqrt{N_S+N_B}}\,,
\label{eq:significance}
\end{equation}
where $N_S$ is the number of signal events and $N_B$ the number of SM background events inside the component's volume.
The number of signal events $N_S=N(x_1,x_2,\sqrt{s},{\cal L})$, given by eq.\ \eqref{eq:nevents}, inside a detector component (with $x_1$ and $x_2$ being its inner and outer radii) is controlled by the production cross section $\sigma_{\nu N}$ and the lifetime $\uptau^{\rm lab}$ of the heavy neutrinos, both of which are dependent on the squared active-sterile mixing $|\theta|^2$ and the heavy neutrino mass $M$. 
Therefore, a sensitivity of $2\sigma$ for the heavy neutrino search via displaced vertices with given mass $M$ can be defined by the value for $|\theta|^2$ that results in a significance larger than 2.
This relation maps the sensitivity of each detector component to the signal onto the heavy neutrino parameter space.

We show a schematic illustration of the discussed mapping of the SiD components into the heavy neutrino parameter space in fig.\ \ref{fig:schematic}. Therein each component is assigned a distinct color, and the order corresponds to the layers of the components inside the SiD. 
Sets of parameters to the right of the inner region lead to a vertex displacement being indistinguishable with the considered vertex resolution, which necessitates a conventional search.
Sets of parameters to the left of the muon system take place dominantly outside the detector and are thus invisible.
The horizontal lines in the figure denote the number of heavy neutrino decays that are to be expected, which scale with $|\theta|^2$ and are proportional to the cross sections (here taken to be flat) shown in fig.\ \ref{fig:Nproduction}.
The vertical dashed line denotes the $W$ boson mass and indicates the limit of this search channel. 
In general, the heavy neutrino mass has to be smaller than the center-of-mass energy of the incident electron-positron beams. However, for $M\sim m_W$ new decay channels for the heavy neutrinos into on-shell $W$ and $Z$ bosons open up, which renders their lifetimes generally too short to allow for a measurable vertex displacement.

For the combined response of the SiD components, see the discussion above in sections \ref{sec:innerregion} to \ref{sec:mucal}, we find:
The search for long-lived heavy neutrinos via displaced vertices is sensitive to displacements as small as $x_{\rm res}$, but only for displacements larger than $10\,\mu$m the search is essentially free of irreducible background. As discussed the detector components are (almost) background-free with the exception of the ECAL. Furthermore, it is unclear if heavy neutrino decays that occur close to the outer radius of the muon identification system are registered.
Since the mapping of the physical extensions of the ECAL and the muon identification system into the heavy neutrino parameter space shows a considerable overlap\footnote{The overlap occurs due to the stochastic nature of the heavy neutrino decays. Therefore the gaps between some of the detector components do not affect the combined sensitivity of the SiD.} with the tracker and the HCAL, we therefore consider vertex displacements between $10\,\mu$m and 249 cm (i.e.\ within the outer radius of the HCAL) as conservative bounds for signal events to be free of background and in principle detectable by the SiD.
Notice that for $M\leq m_W$ the heavy neutrino has a relativistic velocity $\beta \gtrsim 0.1$, such that it decays in the calorimetric system within $\sim 30$ ns after the interaction, which may be important when a trigger is used at circular colliders.

We remark that it is very important to include the muon identification system as an independent search for displaced vertices from heavy neutrino decays at future lepton colliders, which may provide independent and complementary information.

\subsection{Resulting sensitivities}
In this section we present the sensitivities of the future lepton colliders to heavy neutrino searches via displaced vertices. The SiD serves as a benchmark detector for all the experiments with the modi operandi of the FCC-ee, CEPC, and ILC from fig.\ \ref{fig:modioperandi}.

According to the discussion in section \ref{sec:SiD}, we take the heavy neutrino decays with a vertex displacement between $10\,\mu$m and 249 cm to be free of background and detectable by the SiD. The absence of SM background implies that the detection of a single event corresponds to the detection of a heavy neutrino signal via displaced vertices with a significance of 1$\sigma$, cf. eq.\ \eqref{eq:significance}.
In the following, we demand at least four signal events in order to establish a signal at $2\sigma$.

\begin{figure}
\begin{minipage}{0.32\textwidth}
\includegraphics[width=1.0\textwidth]{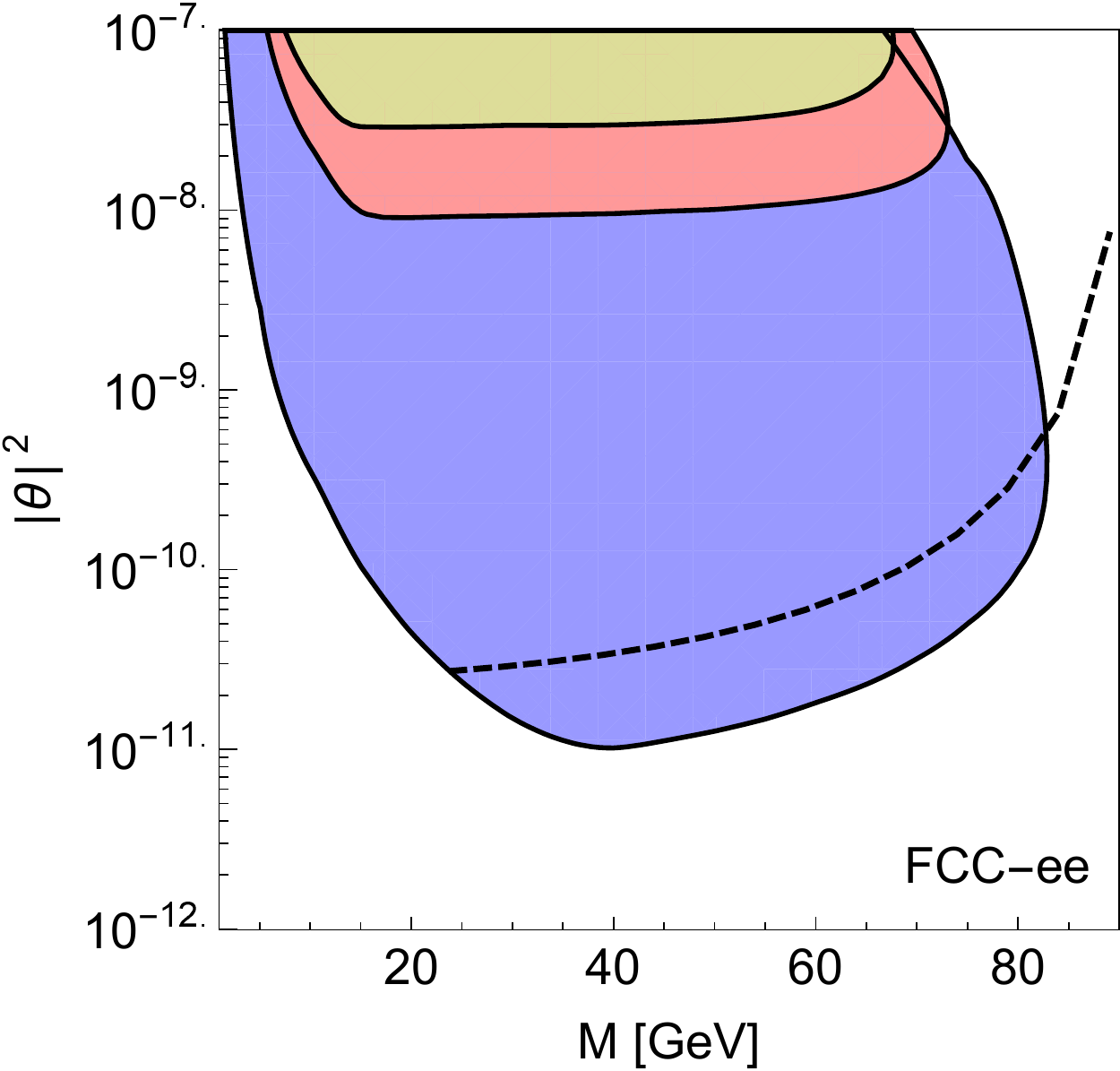}
\end{minipage}
\begin{minipage}{0.32\textwidth}
\includegraphics[width=1.0\textwidth]{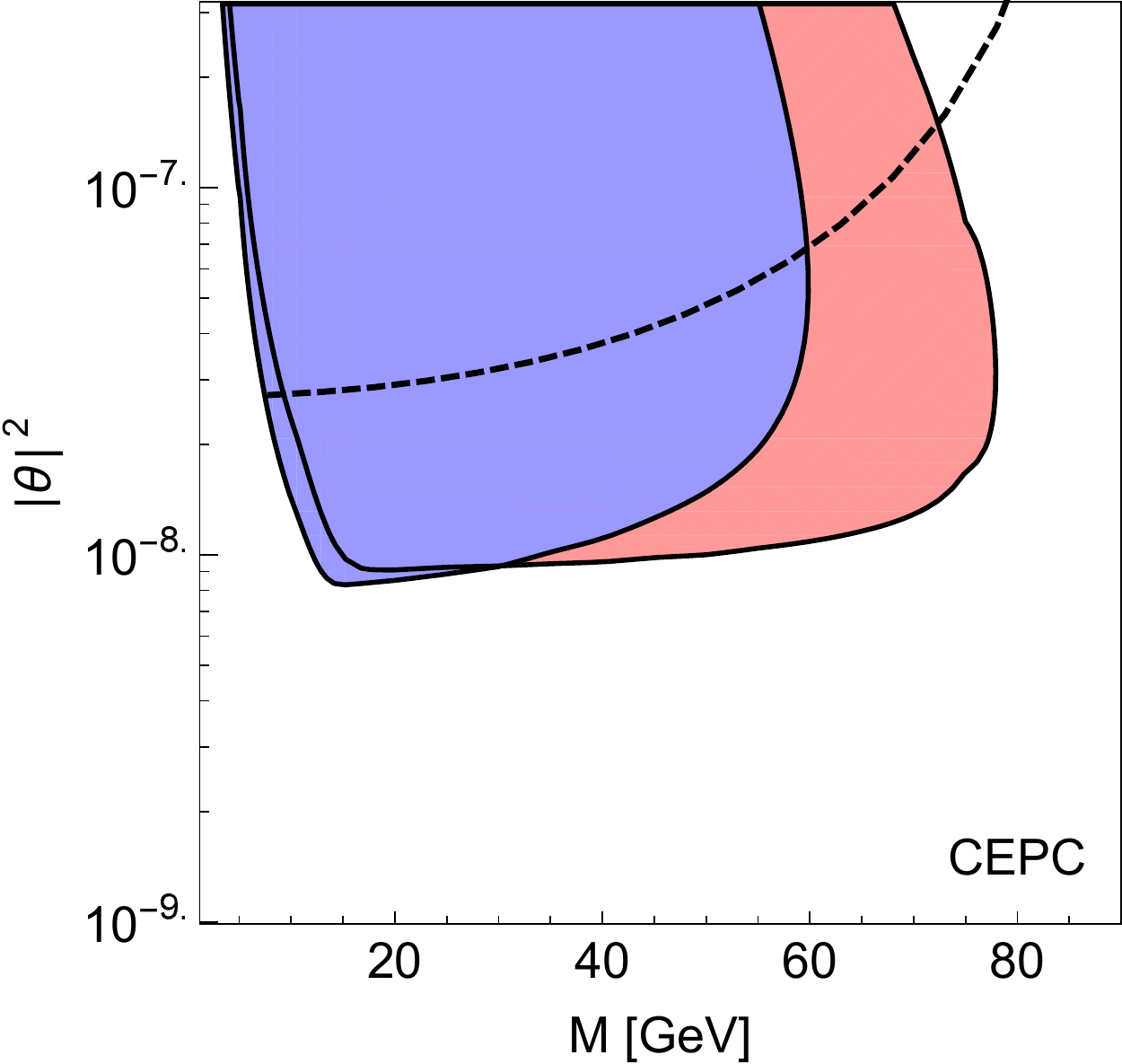}
\end{minipage}
\begin{minipage}{0.32\textwidth}
\includegraphics[width=1.0\textwidth]{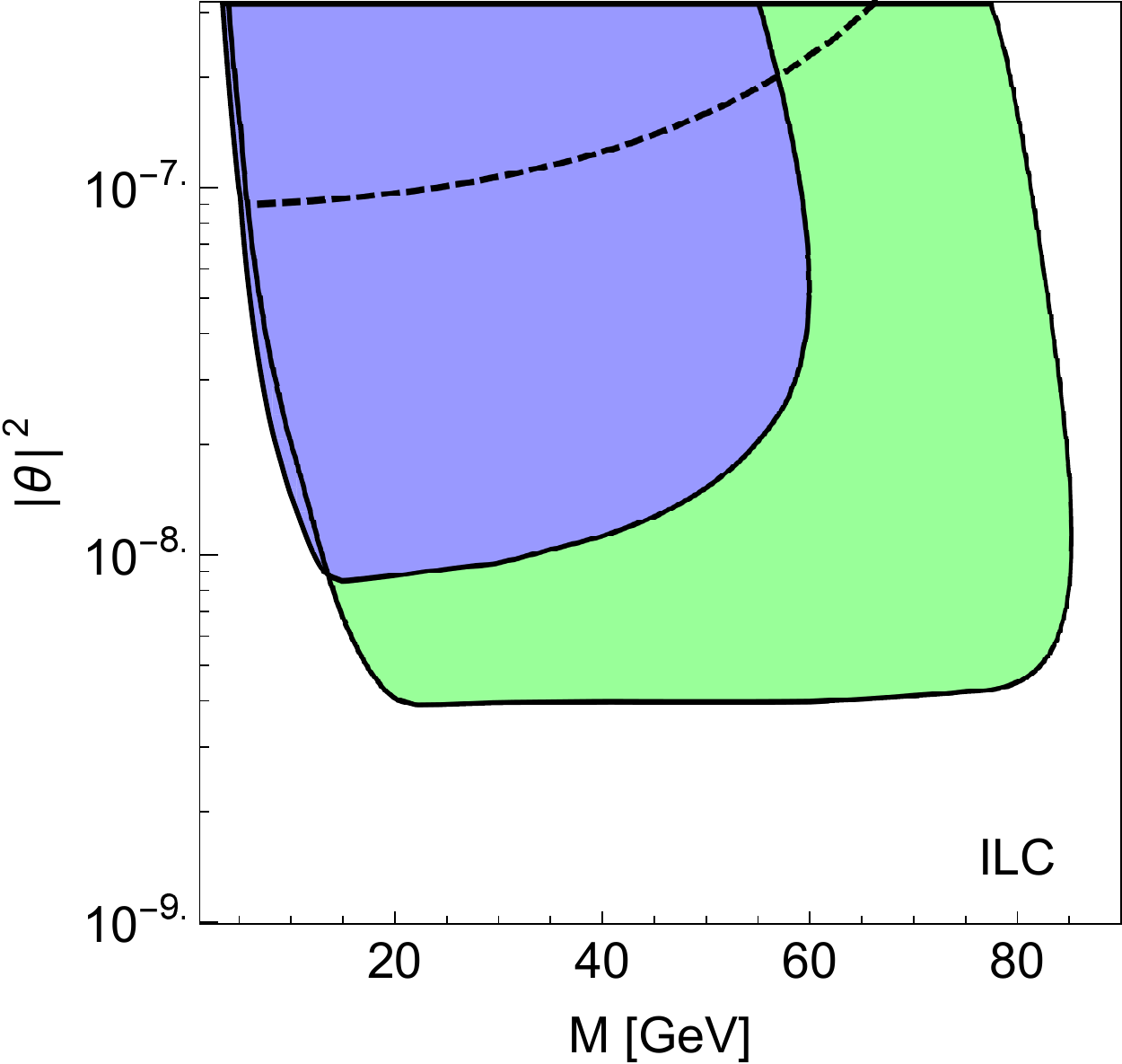}
\end{minipage}

\centering
\includegraphics[width=0.9\textwidth]{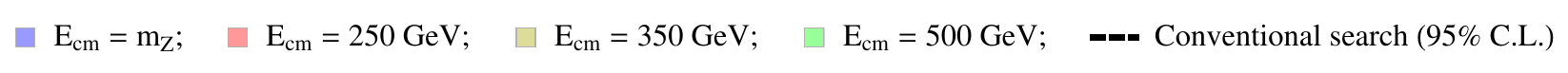}

\caption{Sensitivity at $2\sigma$ for sterile neutrino searches via displaced vertices at the FCC-ee, the CEPC, and the ILC, assuming 100\% signal efficiency.
The colors denote the different modi operandi from fig.\ \ref{fig:modioperandi}. The sensitivities for $E_{cm}\neq m_Z$ are understood for $|\theta|^2 = |\theta_e|^2$ (and $\theta_\mu,\,\theta_\tau=0$).
The SiD is used as benchmark detector for all the lepton collider experiments, for which we found heavy neutrino signals with vertex displacements between 10 $\mu$m and $249$ cm to be essentially free of irreducible background, cf.\ section \ref{sec:SiD}.
The black dashed lines denote the conventional $Z$ pole searches (cf.\ \cite{Antusch:2015mia}).}
\label{fig:results}
\end{figure}

We show the resulting sensitivities of the FCC-ee, the CEPC and the ILC, respectively, to the searches for heavy neutrinos via displaced vertices in fig.\ \ref{fig:results}. Parameter sets of masses and active-sterile mixings inside the colored areas lead to at least four events inside the SiD.
The overall shape of the colored areas can be understood from the schematic illustration in fig.\ \ref{fig:schematic}. We checked that including the muon identification system does not significantly affect the resulting sensitivities for any of the here considered future lepton colliders and their modi operandi. 
For comparison we show estimates for the future sensitivity of the conventional searches at 95\% confidence level by the black, dashed line \cite{Antusch:2015mia}.
This estimate was obtained by a rescaling of the 95\% C.L. exclusion limit from DELPHI with the $Z$ pole luminosities of the respective future lepton collider.

The left-hand plot in fig.\ \ref{fig:results} shows that the $Z$ pole run of the FCC-ee yields the highest sensitivity due to the large envisaged integrated luminosity. This run is sensitive to smaller active-sterile neutrino mixing compared to the estimates for the conventional searches. The physics runs at higher center-of-mass energies show weaker sensitivities when compared to the $Z$ pole run, but they still improve the projected sensitivity of the LHC, which reaches $|\theta|^2 \sim 10^{-7}$ for heavy neutrino masses $\sim 20$ GeV, cf.\ refs.\ \cite{ConstraintsLHC}.

At the CEPC, the considered modi operandi result in the sensitivities for the Higgs run at 250 GeV and the $Z$ pole run to be comparable, with the former being sensitive to larger heavy neutrino masses. 
It is interesting to note that, despite the heavy neutrino production cross sections being more than one order of magnitude smaller compared to the $Z$ pole run (see fig.\ \ref{fig:Nproduction}), also the Higgs run constitutes a feasible search channel for sterile neutrinos via displaced vertex searches due to the considered integrated luminosities.
We remark that the here shown sensitivities for $\sqrt{s}\neq m_Z$ are strictly valid only for $\theta_\mu,\,\theta_\tau=0$ and $|\theta|^2 = |\theta_e|^2$.

The sensitivities for the ILC show that the high-energy run at 500 GeV has a much higher sensitivity compared to the $Z$ pole searches, which is, in analogy to the CEPC, due to the considered integrated luminosities.
Comparing the Higgs run of the CEPC with the 500 GeV run at the ILC, which both consider the same integrated luminosity, we find that the ILC outperforms the CEPC, due to the larger heavy-neutrino-production cross section with beam polarisation.
On the other hand, a significant enhancement of the sensitivities of the CEPC and the ILC at the $Z$ pole run could be achieved when the run times are prolonged.

\section{Summary and Conclusions}
\label{sec:conclusions}
In this work, we have investigated the sensitivity to sterile neutrinos with electroweak-scale Majorana masses from the search for displaced vertices at future lepton colliders.

We deepened and extended previous work on displaced vertex searches for sterile neutrinos at future lepton colliders in various ways:
We considered an explicit low scale seesaw benchmark model, the SPSS, and calculated the heavy-neutrino-production cross section with WHIZARD, including initial state radiation and initial state polarisation (where applicable).\footnote{The SPSS is chosen because it is representative for a wide class of low scale seesaw models, and we therefore expect our results to hold (at least approximately) in a more general context.}
As future lepton colliders, we considered the FCC-ee, CEPC, and the ILC and included the different center-of-mass energies planned for the respective physics programs, i.e.\ the $Z$ pole run, the Higgs run at 240 or 250 GeV, the top threshold scan at 350 GeV and, for the ILC, also 500 GeV.
For a realistic assessment of the sensitivity, we used the ILC's SiD as benchmark detector and put emphasis on its response to the displaced heavy neutrino signal and the conceivable SM backgrounds.
We find that the SiD is sensitive to the signal in an essentially background-free environment (after suitable cuts), for vertex displacements ranging from 10 $\mu$m to the outer radius of the HCAL.
We expect that removing the backgrounds (cf.\ section \ref{sec:SiD}) with suitable cuts will somewhat reduce the signal efficiency. For instance the DELPHI experiment at LEP quotes a signal efficiency of $\sim 25\%$, which would roughly speaking  shift up the maximal sensitivity by a factor of two. However, the efficiency may be higher at a future detector, 
closer to the here assumed 100\% signal efficiency.
We note that for assessing a more realistic number for the signal efficiency, and also for a better understanding of the response and complementarity of the ECAL, HCAL and muon identification system to the heavy neutrino decays within the respective component, a full simulation of the detector acceptance would be desirable.

The resulting sensitivities of sterile neutrino searches via displaced vertices at future lepton colliders are summarized in figure \ref{fig:results}, for a confidence level of $2\sigma$. 
We find that the FCC-ee at the $Z$ pole run with 110~ab$^{-1}$ yields the best sensitivity, down to squared active-sterile mixings as small as $|\theta|^2 \sim 10^{-11}$. Comparing this estimated sensitivity to the one for a conventional search for sterile neutrinos at the $Z$ pole, the displaced vertex search is sensitive to significantly smaller active-sterile mixing angles. 
It turns out that the center-of-mass energies higher than the $Z$ boson mass can already improve the present exclusion limits of the LHC and its projected sensitivities of $|\theta|^2\sim 10^{-7}$ for 300 fb$^{-1}$. 
For the CEPC, the $Z$ pole run and the higher energy run (at 250 GeV) result in comparable sensitivities, while for the ILC the high-energy run (at 500 GeV in the G-20 physics program) results in its best sensitivity.

In summary, our analysis demonstrates that all the modi operandi of all the future lepton colliders can improve the present bounds and the projected LHC reach.
Highest sensitivities to sterile neutrinos can be reached in the mass range between $\sim$ 10 and 80 GeV.
This is complementary to the experiments like SHiP \cite{Alekhin:2015byh}, which has peak sensitivities at lower masses, around 1 GeV.
We thus conclude that the search for displaced vertices at future lepton colliders constitutes a powerful search channel for heavy neutrinos with masses below the $W$ boson mass.

\subsection*{Acknowledgements}
This work has been supported by the Swiss National Science Foundation. We thank M. Ruan, S. Ganjour, C. Potter, and M. Dam for valuable discussions and feedback. We are indebted to M. Stanitzki for invaluable discussions on lepton colliders and detector concepts in general, and the ILC and SiD in particular.

\bibliographystyle{unsrt}

\end{document}